\documentclass[11pt]{article} 

\usepackage[top=60pt,bottom=70pt,left=75pt,right=75pt]{geometry}

\usepackage{amsmath, amssymb, amsfonts, latexsym, euscript, mathalfa, mathrsfs}
\usepackage{graphicx, color, epsfig, xcolor, subfigure}

\usepackage{setspace, sectsty}

\usepackage{braket, cancel}
\usepackage{tensor}

\usepackage{jheppub}

\newcommand{\nn}{\nonumber}

\newcommand{\e}{\mathrm{e}}
\newcommand{\ii}{\text{i}}

\begin{document}

\baselineskip=18pt  
\numberwithin{equation}{section}  
\allowdisplaybreaks  

\begin{titlepage}

\begin{center}
 {\Large\bf Superstrings on Penrose limits of AdS$_7$ solutions}
\end{center}

\begin{center}

Achilleas Passias\footnote{achilleas.passias@lpthe.jussieu.fr}

\vspace{5mm}

Sorbonne Universit\'{e}, UPMC Paris 06, UMR 7589, LPTHE, \\ 75005 Paris, France  

\vspace*{.5cm}

{\bf Abstract}

\end{center}

{\setlength{\parindent}{0cm}
We consider the Penrose limit  of AdS$_7$ solutions of massive Type IIA supergravity.
The resulting pp-wave geometry is supported by RR and NSNS fields. We quantize the Green--Schwarz superstring on the obtained pp-wave background, in the light-cone gauge.
}

\end{titlepage}

\tableofcontents

\section{Introduction}
The presence of Ramond--Ramond fields poses a technical challenge to the quantization of superstrings. An exception are pp-wave spacetimes which provide curved backgrounds with Ramond--Ramond fields where superstrings can be quantized. String theory on pp-wave spacetimes is of interest in the context of the AdS/CFT correspondence, as pp-wave spacetimes arise in the Penrose limit of anti-de Sitter solutions. In \cite{Berenstein:2002jq} it was proposed that the Penrose limit of the Type IIB AdS$_5 \times S^5$ solution corresponds to singling out a subset of operators in the dual $\mathcal{N}=4$ supersymmetric U$(N)$ Yang--Mills theory, where both the conformal dimension $\Delta$ and a  U$(1)$ R-charge $J$ are taken to be of order $\sqrt{N}$ with the difference $\Delta - J$ finite as $N \to \infty$. 

In this paper we consider the Penrose limit of AdS$_7$ solutions of (massive) Type IIA supergravity, which are supported by RR and NSNS fluxes. These are dual to six-dimensional $\mathcal{N}=(1,0)$ superconformal field theories which admit a description in terms of linear quivers. Upon obtaining the pp-wave spacetime in the Penrose limit, we quantize the Green--Schwarz closed superstring in the light-cone gauge. The Hamiltonian has a ``harmonic oscillator'' form. 
The identification of the operators of the dual six-dimensional field theories which correspond to the string excitations is under consideration.

The rest of the paper is organised as follows: in section \ref{AdS(7)} we review the family of AdS$_7$ solutions and select two of them on which we focus on the remainder of the paper. In section \ref{PenroseLimit} we obtain the pp-wave background in the Penrose limit and analyze its supersymmetry. In section \ref{SuperstringQuantization} we perform the quantization of the Green--Schwarz closed superstring in the light-cone gauge.

\section{AdS$_7$ solutions}\label{AdS(7)}
The AdS$_7$ solutions of massive Type IIA supergravity \cite{Apruzzi:2013yva, Apruzzi:2015wna}, in the form introduced in \cite{Cremonesi:2015bld}, are
	\begin{align}\label{AdS7}
		&\frac1{\pi \sqrt2} ds^2= 8\sqrt{-\frac \alpha{\ddot \alpha}}ds^2({\rm AdS}_7)+ \sqrt{-\frac {\ddot \alpha}\alpha} \left(dz^2 + \frac{\alpha^2}{\dot \alpha^2 - 2 \alpha \ddot \alpha} ds^2(S^2)\right)\,; \nn \\
		&B=\pi \left( -z+\frac{\alpha \dot \alpha}{\dot \alpha^2-2 \alpha \ddot \alpha}\right) {\rm vol}(S^2)\,, \quad e^\phi=2^{5/4}\pi^{5/2} 3^4 \frac{(-\alpha/\ddot \alpha)^{3/4}}{(\dot \alpha^2-2 \alpha \ddot \alpha)^{1/2}}\,; \nn \\ 
		&F_2 =  \left(\frac{\ddot \alpha}{162 \pi^2}+ \frac{\pi F_0\alpha \dot \alpha}{\dot \alpha^2-2 \alpha \ddot \alpha}\right) {\rm vol}(S^2)\,, \quad F_0\,.
	\end{align}	

They are determined by a cubic in $z$ function $\alpha(z)$ and its derivatives (a dot ($\dot{}$) denotes derivation with respect to $z$) which is subject to:
\begin{equation}
\dddot{\alpha} = -162 \pi^3 F_0\,.
\end{equation}

$B$ is the NSNS two-form potential, $\phi$ the dilaton, $F_2$ the RR two-form field strength and $F_0$ the Romans mass.

We will endow the seven-dimensional anti-de Sitter spacetime AdS$_7$ and the two-sphere $S^2$ with the metrics
\begin{equation}
ds^2({\rm AdS}_7) = -\cosh^2\rho dt^2 + d\rho^2 + \sinh^2\rho ds^2(S^5)\,, \quad
ds^2(S^2) = d\theta^2 + \sin^2\theta d\psi^2\,,
\end{equation}
where $ds^2(S^5)$ is the unit-radius metric on the five-sphere $S^5$. Also, in \eqref{AdS7}, ${\rm vol}(S^2) := \sin\theta d\theta \wedge d\psi$.

\begin{figure}[ht]
\centering	
    \subfigure[\label{fig:R3-D6}]{\includegraphics[width=6cm]{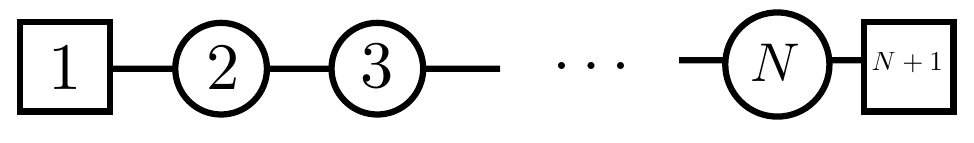}}
\hspace{2cm}
	\subfigure[\label{fig:massless}]{\includegraphics[width=6cm]{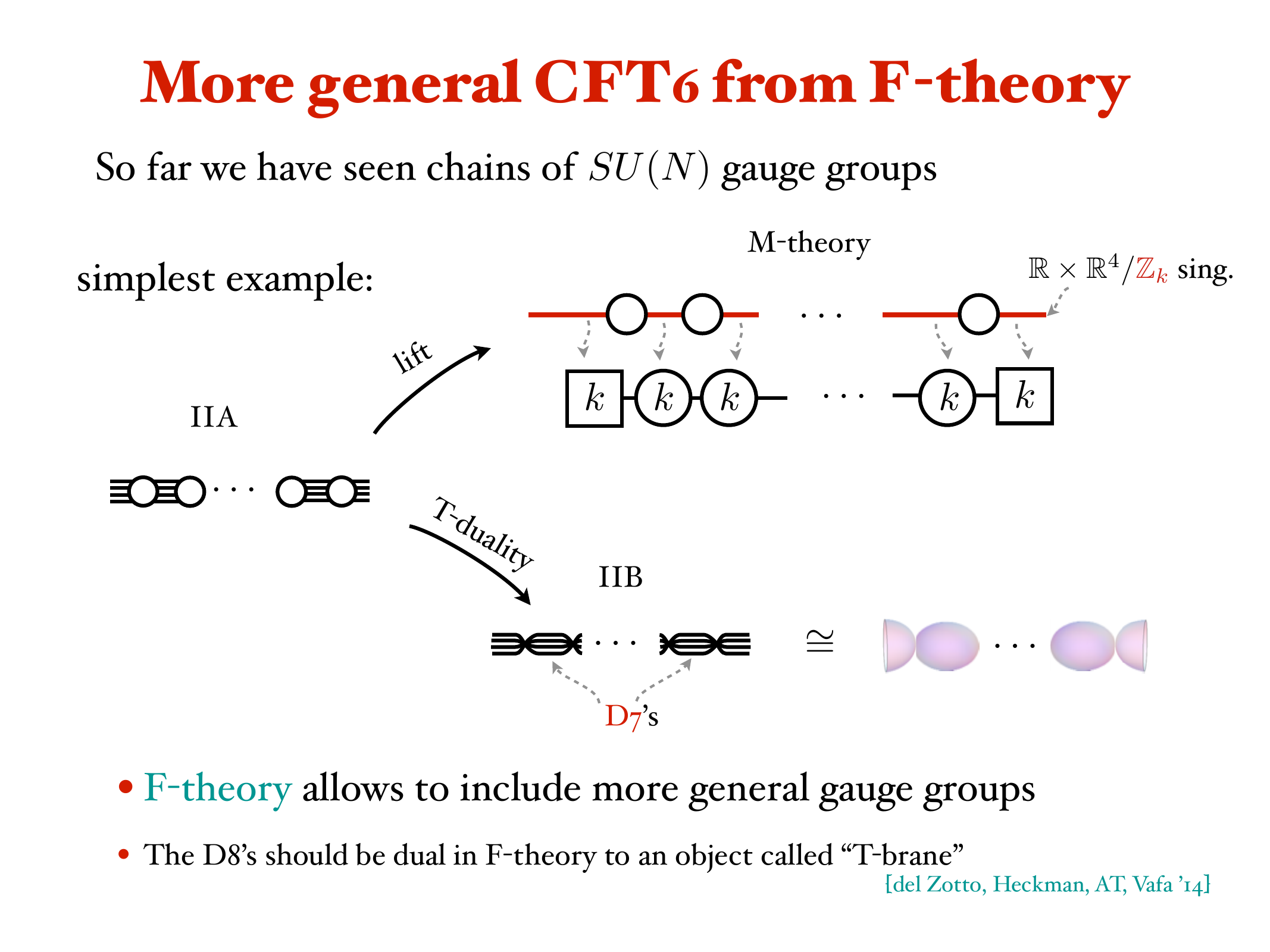}}
	\caption{Linear quivers for the six-dimensional $\mathcal{N}=(1,0)$ field theories dual to the Type IIA supergravity solutions (a) and (b).}
	\label{fig:cft}
\end{figure}

We will consider two representative solutions (a) and (b). 

Solution (a) is determined by
\begin{equation}
\alpha_{\rm (a)}(z) = \frac{27}{2} n_0 \pi^2 z (N^2-z^2)\,,
\end{equation}
where $n_0 \in \mathbb{Z}$ are the quanta of the Romans mass and $N \in \mathbb{Z}$ the NSNS flux quanta. $z \in [0,N]$; at $z=0$ the geometry is regular while at $z=N$ there is a D6-brane type singularity. The dual field theory is described by the quiver in Figure \ref{fig:R3-D6}.

Solution (b) has zero Romans mass and is determined by
\begin{equation}
\alpha_{\rm (b)}(z) = \frac{81}{2} k \pi^2 z (N-z)\,,
\end{equation}
where $k \in \mathbb{Z}$ are the RR flux quanta and $N \in \mathbb{Z}$ the NSNS flux quanta. $z \in [0,N]$, and at the endpoints of the interval there are D6-brane type singularities. The dual field theory is described by the quiver in Figure \ref{fig:massless}.

\section{The Penrose limit}\label{PenroseLimit}
In order to define the Penrose limit, we consider null geodesic motion in the $\psi$ (isometry) direction with afffine parameter $\lambda$. Furthermore, we will take the geodesic located at the center of AdS$_7$ at $\rho=0$ and at the equator of the $S^2$ at $\theta = \pi/2$. Thus, the U$(1) \subset$ SU$(2)$ isometry of the two-sphere is preserved, and similarly for the isometries of the five-sphere in AdS$_7$. 

The geodesic equation 
\begin{equation}
\frac{d^2 x^\mu}{d\lambda^2} + 
\Gamma^\mu_{\rho\sigma} \frac{dx^\rho}{d\lambda} \frac{dx^\sigma}{d\lambda} = 0\,,
\end{equation}
yields
\begin{equation}
t'' = 0, \qquad \psi'' = 0, \qquad g^{zz}(g_{\psi\psi,z} (\psi')^2 + g_{tt,z} (t')^2) = 0\,,
\end{equation}
while imposing $g_{\mu\nu} (x^\mu)' (x^\nu)' = 0$ (null condition) yields
\begin{equation}
g_{\psi\psi} (\psi')^2 + g_{tt} (t')^2 = 0\,.
\end{equation}
Substituting in the last equation coming from the geodesic equation we have
\begin{equation}
g^{zz}(- g_{\psi\psi,z} \frac{g_{tt}}{g_{\psi\psi}} + g_{tt,z})(t')^2 = 0\,,
\end{equation}
which yields 
\begin{equation}\label{condition}
\frac{\dot{\alpha}\left(\ddot\alpha(\dot \alpha^2 - 2 \alpha \ddot \alpha) + \alpha \dot{\alpha}\dddot{\alpha}\right)}{\ddot\alpha^2(\dot \alpha^2 - 2 \alpha \ddot \alpha)} = 0\,.
\end{equation}
We conclude that $z$ is fixed at a critical value $z_c$ for which \eqref{condition} is satisfied. For the specific solutions we will study we will see that $z_c$ is such that $\dot{\alpha}(z_c) = 0$.

For solution (a) we have
\begin{equation}
z_c = \frac{N}{\sqrt{3}}
\end{equation} 
and, as mentioned, $\dot{\alpha}(z_c) = 0$.

We introduce
\begin{equation}
\zeta := \frac{z}{z_c}-1
\end{equation}
and expand the metric near $\zeta = 0$:
\begin{align}
L^{-2} ds^2 = \left(1 - \frac{1}{2} \zeta - \frac{3}{8} \zeta^2 \right)ds^2({\rm AdS}_7) 
+ \frac{3}{8} d\zeta^2 
+ \frac{1}{16} \left(1 - \frac{1}{2} \zeta - \frac{15}{8} \zeta^2 \right)ds^2(S^2) + \mathcal{O}(\zeta^3)\,, 
\end{align}	
where
\begin{equation}
L^2 := \frac{8\sqrt{2}\pi}{3} N \,.
\end{equation}
We now make the coordinate transformation
\begin{equation}
t = x^+ + \frac{x^-}{L^2}\,, \quad \psi = 4x^+ - \frac{4x^-}{L^2}\,, \quad
\rho = \frac{r}{L}\,, \quad \theta = \pi/2 + \frac{4x^7}{L}\,, \quad 
\zeta = \sqrt{\frac{8}{3}}\frac{x^8}{L}\,.
\end{equation}
The Penrose limit is $L \to \infty$ i.e.\ $N \to \infty$. The metric becomes:
\begin{equation}\label{ppwave}
ds^2 = - 4 dx^+ dx^- - \sum_{I=1}^8 m_I^2 (x^I)^2(dx^+)^2 + dr^2 + r^2 ds^2(S^5) + (dx^7)^2 + (dx^8)^2
\end{equation}
where 
\begin{equation}
(m_1,m_2,m_3,m_4,m_5,m_6,m_7,m_8) := (1,1,1,1,1,1,4,2)\,.
\end{equation}

In the limit under consideration, the dilaton is
\begin{equation}
e^{\phi} = \frac{2^{3/4} \sqrt{\pi}}{N^{1/2} n_0}\,.
\end{equation}
In order to have a finite dilaton we need to take $n_0 \to 0$ ($F_0 \to 0$) such that $N^{1/2} n_0$ stays finite as $N \to \infty$.

Following the same procedure for the rest of the fields we obtain
\begin{align}\label{ppwavefields}
B = -6 x^8 dx^+ \wedge dx^7\,, \quad
F_2 = - 4 e^{-\phi} dx^+ \wedge dx^7\,, \quad
F_0 = 0\,.
\end{align}

For solution (b) we have:
\begin{equation}
z_c = \frac{N}{2}\,.
\end{equation}

We introduce again
\begin{equation}
\zeta := \frac{z}{z_c}-1
\end{equation}
and expand the metric near $\zeta = 0$:
\begin{align}
L^{-2} ds^2 = \left(1 - \frac{1}{2} \zeta^2 \right)ds^2({\rm AdS}_7) 
+ \frac{1}{4} d\zeta^2 
+ \frac{1}{16} \left(1 - \frac{3}{2} \zeta^2 \right)ds^2(S^2) + \mathcal{O}(\zeta^3) \,,
\end{align}	
where
\begin{equation}
L^2 := 4 \pi N\,.
\end{equation}
We now make the coordinate transformation
\begin{equation}
t = x^+ + \frac{x^-}{L^2}\,, \quad \psi = 4x^+ - \frac{4x^-}{L^2}\,, \quad
\rho = \frac{r}{L}\,, \quad \theta = \pi/2 + \frac{4x^7}{L}\,, \quad 
\zeta = 2 \frac{x^8}{L}\,.
\end{equation}
The Penrose limit is $L \to \infty$ i.e.\ $N \to \infty$. The metric becomes that of
\eqref{ppwave}.

In the limit under consideration, the dilaton is
\begin{equation}
e^{\phi} = \frac{N^{1/2}\sqrt{\pi}}{k}\,.
\end{equation}
In order to have a finite dilaton we need to take $k \to \infty$ such that $\frac{N^{1/2}}{k}$ stays finite as $N \to \infty$.

For the rest of the fields we find the same expressions as \eqref{ppwavefields}.

The pp-wave backgound for both solutions (a) and (b) is the same except for the constant value of the dilaton. It is straightforward to check that the equations of motion are satisfied. For instance, the non-trivial component of the Einstein equations is:\footnote{Given a $p$-form $C$, we use the notation:
\begin{equation}
C^2:= C_{M_1\dots M_p} C^{M_1\dots M_p}\,, \quad
(C^2)_{MN} := C_{MM_1\dots M_{p-1}} C_N{}^{M_1\dots M_{p-1}}\,.
\end{equation}
}
\begin{equation}\label{einstein}
R_{++} + 2\nabla_+\nabla_+ \phi - \frac{1}{4} (H^2)_{++} - e^{2\phi}\left[\frac{1}{2}(F^2_2)_{++} - \frac{1}{8} g_{++} F_2^2 \right]= 0 \,,
\end{equation}
where $R_{++}$ is the only non-zero component of the Ricci tensor. Given that
\begin{equation}
R_{++} = 26\,, \quad 
\nabla_+\nabla_+ \phi = 0\,, \quad
(H^2)_{++} = 72\,, \quad 
(F^2_2)_{++} = 16 e^{-2\phi}\,, \quad
F_2^2 = 0\,,
\end{equation}
\eqref{einstein} is satisfied.

We can express the light-cone momentum in terms of the conformal dimension $\Delta$ and the R-charge $J$ of a dual operator \cite{Berenstein:2002jq}:
\begin{subequations}
\begin{align}
2 p^- &= \ii \partial_{x^+} = \ii (\partial_t + 4\partial_\psi) = \Delta - 4 J \,, \\
2 p^+ &= \ii \partial_{x^-} = \frac{\ii }{L^2} (\partial_t - 4\partial_\psi) = \frac{1}{L^2}(\Delta + 4 J)\,.
\end{align}
\end{subequations}

\subsection{Supersymmetry}
We introduce the tangent space frame
\begin{equation}
\e^+ = dx^+\,, \quad 
\e^- = dx^- + \frac{1}{4} \sum_{I=1}^{8} m^2_I (x^I)^2 dx^+\,, \quad 
\e^I = dx^I\,, 
\end{equation}
$I=1,2,\dots,8$ and tangent space metric $\eta$ with components
\begin{equation}
\eta_{++}=\eta_{--}= 0\,, \quad
\eta_{+-}=\eta_{-+}=-2\,, \quad 
\eta_{IJ} = \delta_{IJ}\,,
\end{equation}
$I,J=1,2,\dots,8$ such that
\begin{equation}
ds^2 = \eta_{AB} \e^A \e^B\,, \quad A,B = 0,1,\dots,9\,,
\end{equation}
and
\begin{equation}
\{\Gamma_A, \Gamma_B \} = 2\eta_{AB}\,,
\end{equation}
for the Clifford algebra generators. We take $\Gamma_\pm = -\Gamma_0 \pm \Gamma_9$.

The equations coming from setting the gravitini supersymmetry variations to zero are:\footnote{We work in the democratic formulation of Type IIA supergravity.}
\begin{align}
0 &= \left(\nabla_M - \frac{1}{4} \cancel{H}_M \right) \epsilon_1 - \frac{1}{16} e^{\phi}  \sum_{k=0}^5 \cancel{F_{2k}} \Gamma_M
\epsilon_2\,, \\
0 &= \left(\nabla_M + \frac{1}{4} \cancel{H}_M \right) \epsilon_2 - \frac{1}{16} e^{\phi} \sum_{k=0}^5 (-1)^{\lfloor k \rfloor} \cancel{F_{2k}} \Gamma_M
 \epsilon_1 \,.
\end{align}
$\epsilon_1$ and $\epsilon_2$ are Majorana--Weyl spinors of positive and negative chirality respectively. 
The equations coming from setting the dilatini supersymmetry variations to zero are:
\begin{align}\label{10dSusyVariationsD}
0 &= \left(\cancel{\partial} \phi - \frac{1}{2} \cancel{H} \right) \epsilon_1 - \frac{1}{16} e^{\phi} \Gamma^M \sum_{k=0}^5 \cancel{F_{2k}} \Gamma_M
\epsilon_2\,, \\
0 &= \left(\cancel{\partial} \phi + \frac{1}{2} \cancel{H} \right) \epsilon_2 - \frac{1}{16} e^{\phi} \Gamma^M \sum_{k=0}^5 (-1)^{\lfloor k \rfloor} \cancel{F_{2k}} \Gamma_M
\epsilon_1\,.   
\end{align}

We now substitute the pp-wave background of the previous section.
The equations coming from the dilatini variations are:
\begin{align}
\Gamma^{+7}(\Gamma^8 \epsilon_1 + \epsilon_2) = 0\,, \quad
\Gamma^{+7}(\Gamma^8 \epsilon_2 + \epsilon_1) = 0\,.
\end{align}
The equations coming from the gravitini variations are:
\begin{subequations}
\begin{align}
&\partial_{x^i} \epsilon_1 + \frac{1}{2} \Gamma^{+7} \Gamma_i \epsilon_2 = 0\,,\quad 
\partial_{x^i} \epsilon_2 - \frac{1}{2} \Gamma^{+7} \Gamma_i \epsilon_1 = 0\,; \\
&\left(\partial_{x^7} - \frac{3}{2} \Gamma^{+8}\right) \epsilon_1 + \frac{1}{2} \Gamma^{+7} \Gamma_7 \epsilon_2 = 0\,, \quad 
\left(\partial_{x^7} + \frac{3}{2} \Gamma^{+8}\right) \epsilon_2 - \frac{1}{2} \Gamma^{+7} \Gamma_7 \epsilon_1 = 0\,; \\
&\left(\partial_{x^8} + \frac{3}{2} \Gamma^{+7}\right) \epsilon_1 + \frac{1}{2} \Gamma^{+7} \Gamma_8 \epsilon_2 = 0\,, \quad 
\left(\partial_{x^8} - \frac{3}{2} \Gamma^{+7}\right) \epsilon_2 - \frac{1}{2} \Gamma^{+7} \Gamma_8 \epsilon_1 = 0\,; \\
&\partial_{x^-} \epsilon_1 + \frac{1}{2} \Gamma^{+7} \Gamma_- \epsilon_2 = 0\,, \quad
\partial_{x^-} \epsilon_2 - \frac{1}{2} \Gamma^{+7} \Gamma_- \epsilon_1 = 0\,; \\
&\left(\partial_{x^+} + \sum_{I=1}^8 \frac{1}{4} m^2_I x^I \Gamma^{I+} + \frac{3}{2} \Gamma^{78} \right) \epsilon_1 + \frac{1}{2} \Gamma^{+7} \Gamma_+ \epsilon_2 = 0\,, \nn \\
&\left(\partial_{x^+} + \sum_{I=1}^8 \frac{1}{4} m^2_I x^I \Gamma^{I+} - \frac{3}{2} \Gamma^{78}\right) \epsilon_2 - \frac{1}{2} \Gamma^{+7} \Gamma_+ \epsilon_1 = 0\,.
\end{align}
\end{subequations}
where we have used that the non-zero components of the spin connection are
\begin{equation}
{\omega_+}_{i+} = x^i\,,    \quad  
{\omega_+}_{7+} = 16 x^7\,, \quad  
{\omega_+}_{8+} = 4 x^8\,.
\end{equation}
We find that are solved by
\begin{equation}
\begin{pmatrix}
\epsilon_1 \\ \epsilon_2 
\end{pmatrix}
= 
\begin{pmatrix}
e^{-\frac{3}{2} \Gamma^{78} x^+} \cos x^+ \chi_1 \\
-e^{+\frac{3}{2} \Gamma^{78} x^+} \sin x^+ \Gamma^7 \chi_1
\end{pmatrix}
\,, \quad
\begin{pmatrix}
\epsilon_1 \\ \epsilon_2 
\end{pmatrix}
= 
\begin{pmatrix}
e^{-\frac{3}{2} \Gamma^{78} x^+} \sin x^+ \Gamma^7 \chi_2 \\
-e^{+\frac{3}{2} \Gamma^{78} x^+} \cos x^+ \chi_2
\end{pmatrix}\,,
\end{equation}
where $\chi_1$, $\chi_2$ are constant Majorana--Weyl spinors of opposite chirality and in addition, $\epsilon_1$, $\epsilon_2$ are subject to
\begin{equation}
\Gamma^+ \epsilon_1 = 0 = \Gamma^+ \epsilon_2\,.
\end{equation}
In total sixteen supersymmetries are preserved, as for the AdS$_7$ solutions.

\section{Light-cone Green--Schwarz string quantization}\label{SuperstringQuantization}

\subsection{Bosonic part}
The bosonic part of the Green--Schwarz action is
\begin{equation}
S_{\rm b} = - \frac{1}{4\pi\alpha'} \int d^2\sigma \sqrt{-h} (h^{ab} \partial_a X^\mu \partial_b X^{\nu} G_{\mu\nu} + \varepsilon^{ab} \partial_a X^\mu \partial_b X^{\nu} B_{\mu\nu})\,.
\end{equation}
We fix:
\begin{equation}\label{h}
h^{\sigma\tau} = 0\,, \quad h^{\sigma\sigma} = - h^{\tau\tau} = 1\,.
\end{equation}
It follows that $X^+$ satisfies the wave equation in two dimensions
\begin{equation}
(-\partial_\tau^2 + \partial_\sigma^2) X^+ = 0\,,
\end{equation}
and so we can further impose the light-cone gauge
\begin{equation}
X^+ = \tau\,.
\end{equation}
The bosonic string action then becomes
\begin{equation}\label{bosonicaction}
S_{\rm b} = - \frac{1}{4\pi\alpha'} \int d^2\sigma \left(4 \partial_\tau X^- - \sum_{I=1}^8 (\partial_\tau X^I)^2 + \sum_{I=1}^8 (\partial_\sigma X^I)^2  + \sum_{I=1}^8 m^2_I (X^I)^2 -12 X^8 \partial_\sigma X^7\right)\,.
\end{equation}

Introducing the center-of-mass variable
\begin{equation}
x^-(\tau) = \frac{1}{\ell} \int_0^\ell d\sigma X^-(\tau, \sigma)\,,
\end{equation}
we have
\begin{equation}
p^+ = - \frac{1}{2} p_- = - \frac{1}{2} \frac{\delta{L_{\rm b}}}{\delta(\partial_\tau x^-)} = \frac{\ell}{2\pi\alpha'}\,,
\end{equation}
where $L_{\rm b}$ is the (bosonic) Lagrangian. Hence,
\begin{equation}
\ell = 2 \pi \alpha' p^+\,.
\end{equation}

We consider closed strings so impose the periodicity condition
\begin{equation}
X^M(\tau, \sigma + 2 \pi \alpha'p^+) = X^M(\tau, \sigma)\,.
\end{equation}

The equations of motion that follow from the action \eqref{bosonicaction} are then:
\begin{align}
-\partial^2_\tau X^i+\partial_\sigma^2 X^i -  X^i &= 0\,, \\
-\partial^2_\tau X^7+\partial_\sigma^2 X^7 -16X^7 - 6\partial_\sigma X^8 &= 0\,, \\
-\partial^2_\tau X^8+\partial_\sigma^2 X^8 - 4X^8 + 6\partial_\sigma X^7 &= 0\,.
\end{align}
where $i=1,2,\dots,6$.

For $i=1,2,\dots,6$ the solution is
\begin{align}
X^i(\sigma,\tau) = x_0^i \cos\tau + \frac{p_0^i}{p^+} \sin\tau 
+ &\sqrt{\frac{\alpha'}{2}} \sum_{n=1}^\infty \frac{1}{\sqrt{\omega^i_n}}
\Bigl(\alpha^i_n e^{-\frac{\ii}{\alpha' p^+}(\omega^i_n \tau + n \sigma)} + 
\tilde{\alpha}^i_n e^{-\frac{\ii}{\alpha' p^+}(\omega^i_n \tau - n \sigma)} 
\nn \\ 
&+
(\alpha^i_n)^\dagger e^{\frac{\ii}{\alpha' p^+}(\omega^i_n \tau + n \sigma)} + 
(\tilde{\alpha}^i_n)^\dagger e^{\frac{\ii}{\alpha' p^+}(\omega^i_n \tau - n \sigma)} 
\Bigr)\,,
\end{align}
where the frequencies
\begin{equation}
\omega^i_n = \omega_n := \sqrt{n^2 + (\alpha' p^+)^2}\,.
\end{equation}

For $\underline{i}=7,8$ the solution is
\begin{equation}
X^{\underline{i}} = x_0^{\underline{i}} \cos(m_{\underline{i}}\tau) + \frac{p_0^{\underline{i}}}{m_{\underline{i}} p^+} \sin(m_{\underline{i}} \tau)
+ \sqrt{\frac{\alpha'}{2}} \sum_{n=1}^\infty \left( A_n^{\underline{i}}(\tau,\sigma) + \tilde{A}_n^{\underline{i}}(\tau,\sigma) \right)\,,
\end{equation}
where
\begin{subequations}
\begin{align}
A_n^{\underline{i}}(\tau,\sigma) &:= \sum_{\underline{j}=7}^8 \frac{1}{\sqrt{\omega^{\underline{j}}_n}} 
\left((\Lambda_n)^{\underline{i}}{}_{\underline{j}} \alpha^{\underline{j}}_n e^{-\frac{\ii}{\alpha' p^+}(\omega^{\underline{j}}_n \tau + n \sigma)} 
+ (\bar{\Lambda}_n)^{\underline{i}}{}_{\underline{j}}(\alpha^{\underline{j}}_n)^\dagger e^{\frac{\ii}{\alpha' p^+}(\omega^{\underline{j}}_n \tau + n \sigma)} \right)\,, \\
\tilde{A}_n^{\underline{i}}(\tau,\sigma) &:= \sum_{\underline{j}=7}^8 \frac{1}{\sqrt{\omega^{\underline{j}}_n}} 
\left((\bar{\Lambda}_n)^{\underline{i}}{}_{\underline{j}} \tilde{\alpha}^{\underline{j}}_n e^{-\frac{\ii}{\alpha' p^+}(\omega^{\underline{j}}_n \tau - n \sigma)} 
+ (\Lambda_n)^{\underline{i}}{}_{\underline{j}}(\tilde{\alpha}^{\underline{j}}_n)^\dagger e^{\frac{\ii}{\alpha' p^+}(\omega^{\underline{j}}_n \tau - n \sigma)} \right)\,,
\end{align}
\end{subequations}
with
\begin{equation}
\begin{pmatrix}
(\Lambda_n)^7{}_7 & (\Lambda_n)^7{}_8 \\ (\Lambda_n)^8{}_7 & (\Lambda_n)^8{}_8
\end{pmatrix}
=
\begin{pmatrix}
\frac{-\ii n}{\sqrt{2}\sqrt{(\alpha' p^+)^2+n^2}\sqrt{1-\frac{\alpha'p^+}{\sqrt{(\alpha' p^+)^2+n^2}}}} &
\frac{\ii n}{\sqrt{2}\sqrt{(\alpha' p^+)^2+n^2}\sqrt{1+\frac{\alpha'p^+}{\sqrt{(\alpha' p^+)^2+n^2}}}}
\\ 
\frac{\sqrt{1-\frac{\alpha'p^+}{\sqrt{(\alpha' p^+)^2+n^2}}}}{\sqrt{2}} &
\frac{\sqrt{1+\frac{\alpha'p^+}{\sqrt{(\alpha' p^+)^2+n^2}}}}{\sqrt{2}}
\end{pmatrix}\,,
\end{equation}
and
\begin{equation}
\omega^7_n :=  \sqrt{n^2 + (\alpha' p^+)^2} +  3 \alpha' p^+\,, \quad
\omega^8_n := |\sqrt{n^2 + (\alpha' p^+)^2} -  3 \alpha' p^+|\,.
\end{equation}

$\Lambda_n$ satisfies
\begin{equation}
\begin{pmatrix}
\frac{(\omega^{\underline{i}}_n)^2 -n^2}{(\alpha' p^+)^2} -16 & \frac{6 \ii n}{\alpha' p^+} \\
-\frac{6 \ii n}{\alpha' p^+} & \frac{(\omega^{\underline{i}}_n)^2 -n^2}{(\alpha' p^+)^2} -4
\end{pmatrix}
\begin{pmatrix}
(\Lambda_n)^7{}_{\underline{i}} \\ (\Lambda_n)^8{}_{\underline{i}}
\end{pmatrix}
=0
\end{equation}
and
\begin{equation}
\sum_{\underline{k}=7}^8 (\Lambda_n)^{\underline{i}}{}_{\underline{k}} (\bar{\Lambda}_n)^{\underline{j}}{}_{\underline{k}} = \delta^{\underline{i}\underline{j}}\,,
\quad
\sum_{\underline{k}=7}^8 (\Lambda_n)^{\underline{k}}{}_{\underline{i}} (\bar{\Lambda}_n)^{\underline{k}}{}_{\underline{j}} = \delta_{\underline{i}\underline{j}}\,.
\end{equation}

The canonical quantization condition
\begin{equation}
[X^I(\sigma,\tau), \Pi^I(\sigma',\tau)] = i \delta^{IJ} \delta(\sigma-\sigma')\,,
\end{equation}
where 
\begin{equation}
\Pi^I = \frac{1}{2\pi\alpha'} \partial_\tau X^I\,,
\end{equation} 
yields
\begin{equation}
[x_0^I,p_0^I] = i \delta^{IJ}\,, \quad
[\alpha^I_n, (\alpha^J_m)^\dagger] = \delta^{IJ} \delta_{nm}\,, \quad 
[\tilde\alpha^I_n, (\tilde\alpha^J_m)^\dagger] = \delta^{IJ} \delta_{nm}\,,
\end{equation}
with the rest of the commutators vanishing.

The bosonic Hamiltonian is
\begin{equation}
H_{\rm b} = \frac{1}{4\pi\alpha'}\int_0^{2\pi\alpha'p^+} d\sigma \left( (\partial_\tau X^I)^2 + (\partial_\sigma X^I)^2 + m_I (X^I)^2 - 12 X^8 \partial_\sigma X^7 \right)\,.
\end{equation}
It has the following expression in terms of the creation and annihilation
operators:
\begin{equation}\label{Hb}
H_{\rm b} = \frac{1}{\alpha' p^+} \sum_{I=1}^8 \left[ 
\omega^I_0 (\alpha_0^I)^\dagger \alpha^I_0 +
\sum_{n=1}^\infty \omega_n^I\left((\alpha_n^I)^\dagger \alpha^I_n + (\tilde{\alpha}_n^I)^\dagger \tilde{\alpha}^I_n \right) + \varepsilon_{\rm b}
\right]\,,
\end{equation}
where we have introduced
\begin{equation}
\alpha_0^I := \frac{1}{\sqrt{2m_Ip^+}}\left(p_0^I + \ii m_I p^+ x_0^I\right)\,, \quad
\omega_0^I := m_I \alpha' p^+\,.
\end{equation}
and the normal-ordering constant
\begin{equation}
\varepsilon_{\rm b} := \frac{1}{\alpha'p_+} \sum_{I=1}^8 \left(\frac{1}{2} \omega_0^I + \sum_{n=1}^\infty \omega_n^I \right)\,.
\end{equation}

\subsection{Fermionic part}
The quadratic fermionic part of the Green--Schwarz action is
\begin{equation}
S_{\rm f} = \frac{i}{2\pi\alpha'} \int d^2\sigma \sqrt{-h}(h^{ab} \delta_{\cal I \cal J}-\varepsilon^{ab} (\sigma_3)_{\cal I \cal J}) \partial_a X^M \bar \theta^{\cal I} \Gamma_M (D_b)^{\cal J}{}_{\cal K} \theta^{\cal K}\,,
\end{equation}
where $\mathcal{I}, \mathcal{J}, \mathcal{K} = 1,2$,
\begin{equation}
(D_b)^{\cal J}{}_{\cal K} := \partial_b \delta^{\cal J}{}_{\cal K} + \partial_b X^M \left(\tfrac{1}{2}\omega_M - \tfrac{1}{4} \cancel{H}_M (\sigma_3)^{\cal J}{}_{\cal K}  - \tfrac{1}{8} e^\phi \cancel{F_2} \Gamma_M \epsilon^{\cal J}{}_{\cal K} \right)\,,
\end{equation}
is the pullback to the worldsheet of the supercovariant connection that appears in the gravitini supersymmetry variation, $\theta^1$, $\theta^2$ are Majorana--Weyl spinors of positive and negative chirality respectively and $\bar \theta^{\cal I} := (\theta^{\cal I})^\dagger \Gamma^0$.

We fix $h^{ab}$ as in \eqref{h}. In the light-cone gauge
\begin{equation}
\Gamma^+ \theta ^{\cal I} = 0\,, \quad X^+ = \tau\,,
\end{equation}
and the action becomes
\begin{equation}
S_{\rm f} = \frac{i}{2\pi\alpha'} \int d^2\sigma \bar \theta^{\cal I} \Gamma_+ (-\delta_{\cal I \cal J}(D_\tau)^{\cal J}{}_{\cal K}-(\sigma_3)_{\cal I \cal J}(D_\sigma)^{\cal J}{}_{\cal K})\theta^{\cal K}\,,
\end{equation}
where
\begin{subequations}
\begin{align}
(D_\tau)^{\cal J}{}_{\cal K} &= \partial_\tau \delta^{\cal J}{}_{\cal K} 
+ \left(\tfrac{3}{2} \Gamma^{78} (\sigma_3)^{\cal J}{}_{\cal K} - \Gamma^{7} \epsilon^{\cal J}{}_{\cal K} \right)\,, \\
(D_\sigma)^{\cal J}{}_{\cal K} &= \partial_\sigma \delta^{\cal J}{}_{\cal K}\,.
\end{align}
\end{subequations}
We thus obtain
\begin{align}
S_{\rm f} = \frac{i}{2\pi\alpha'} \int d^2\sigma \Bigl(&-\bar \theta^1 \Gamma_+ (\partial_\tau+\partial_\sigma) \theta^1-\bar \theta^2 \Gamma_+ (\partial_\tau-\partial_\sigma) \theta^2  \nonumber \\ 
&- \frac{3}{2}(\bar \theta^1 \Gamma_+ \Gamma^{78} \theta^1 - \bar \theta^2 \Gamma_+ \Gamma^{78} \theta^2)  
+ \bar \theta^1 \Gamma_+ \Gamma^7 \theta^2 - \bar\theta^2 \Gamma_+ \Gamma^7 \theta^1 \Bigr)\,.
\end{align}

The equations of motion are:
\begin{align}
\left(\partial_\tau + \partial_\sigma + \frac{3}{2} \Gamma^{78} \right) \theta^1 - \Gamma^7 \theta^2 &= 0\,, \\
\left(\partial_\tau - \partial_\sigma - \frac{3}{2} \Gamma^{78} \right) \theta^2 + \Gamma^7 \theta^1 &= 0\,.
\end{align}

We derive the solutions:
\begin{align}
\theta^1 &= e^{-\frac{3}{2} \Gamma^{78}\tau}\frac{1}{2\sqrt{p^+}}(\cos\tau \vartheta^1_0 + \sin\tau \Gamma^7 \vartheta^2_0) + e^{-\frac{3}{2} \Gamma^{78}\tau}\frac{1}{2\sqrt{p^+}}\sum_{n=1}^{\infty} c_n 
\Bigl(
\vartheta^1_n e^{-\frac{\ii}{\alpha' p^+}(\omega_n \tau + n \sigma)} \nn \\ 
&+ \frac{\ii}{\alpha'p^+}\left(\omega_n + n\right)\Gamma^7 \vartheta^2_n e^{-\frac{\ii}{\alpha' p^+}(\omega_n \tau - n \sigma)}
+(\vartheta^1_n)^\dagger e^{\frac{\ii}{\alpha' p^+}(\omega_n \tau + n \sigma)}
- \frac{\ii}{\alpha'p^+}\left(\omega_n + n\right)\Gamma^7 (\vartheta^2_n)^\dagger e^{\frac{\ii}{\alpha' p^+}(\omega_n \tau - n \sigma)}
\Bigr)\,, \\
\theta^2 &= e^{+\frac{3}{2} \Gamma^{78}\tau}\frac{1}{2\sqrt{p^+}}(\cos\tau \vartheta^2_0 -  \sin\tau \Gamma^7 \vartheta^1_0) + e^{+\frac{3}{2} \Gamma^{78}\tau}\frac{1}{2\sqrt{p^+}}\sum_{n=1}^{\infty} c_n \Bigl(\vartheta^2_n e^{-\frac{\ii}{\alpha' p^+}(\omega_n \tau - n \sigma)} \nn \\
&- \frac{\ii}{\alpha'p^+}\left(\omega_n + n\right)\Gamma^7 \vartheta^1_n e^{-\frac{\ii}{\alpha' p^+}(\omega_n \tau + n \sigma)}
+(\vartheta^2_n)^\dagger e^{\frac{\ii}{\alpha' p^+}(\omega_n \tau - n \sigma)}
+ \frac{\ii}{\alpha'p^+}\left(\omega_n + n\right)\Gamma^7 (\vartheta^1_n)^\dagger e^{\frac{\ii}{\alpha' p^+}(\omega_n \tau + n \sigma)}
\Bigr)\,, 
\end{align}
where $\omega_n :=  \sqrt{n^2+(\alpha' p^+)^2}$ and
\begin{equation}
c_n := \sqrt{\frac{1}{1+\frac{1}{(\alpha' p^+)^2}(\omega_n +n)^2}}\,.
\end{equation}

The conjugate momenta are
\begin{equation}
\pi^{\cal I} 
=-\frac{\ii}{2\pi\alpha'} (\theta^{\cal I})^\dagger \Gamma^0 \Gamma_+
=\frac{\ii}{\pi\alpha'} (\theta^{\cal I})^\dagger\,,
\end{equation}
where we have rewritten $\Gamma^0 \Gamma_+ = 2\{\Gamma^+,\Gamma^-\} - 2\Gamma^-\Gamma^+ = -2 - 2\Gamma^-\Gamma^+ $ and used $(\theta^{\cal I})^\dagger \Gamma^- = 0$, which follows from the light-cone gauge condition.

The canonical quantization conditions \cite{Sugiyama:2002tf, Metsaev:2001bj}
\begin{equation}
\{\theta^{1}(\sigma,\tau), \pi^{1}(\sigma',\tau)\} = \frac{\ii}{2} P_+ \delta(\sigma-\sigma')\,, \quad
\{\theta^{2}(\sigma,\tau), \pi^{2}(\sigma',\tau)\} = \frac{\ii}{2} P_- \delta(\sigma-\sigma')\,,
\end{equation}
where
\begin{equation}
P_\pm := \frac{1}{2}(1\pm\Gamma_{11})(-\Gamma^+\Gamma^-)\,,
\end{equation}
yield
\begin{equation}
\{\vartheta^1_n, (\vartheta^1_m)^\dagger\} = P_+ \delta_{nm}\,, \quad
\{\vartheta^2_n, (\vartheta^2_m)^\dagger\} = P_- \delta_{nm}\,.
\end{equation}
The projectors $P_\pm$ reflect the fact that $\theta^{\cal I}$ are Weyl and subject to the light-cone gauge condition.

Upon using the equations of motion the Hamiltonian is given by
\begin{equation}
H_{\rm f} = \frac{\ii}{\pi\alpha'} \int_0^{2\pi\alpha' p^+} d\sigma \Bigl((\theta^1)^\dagger \partial_\tau\theta^1 + (\theta^2)^\dagger \partial_\tau \theta^2 \Bigr)\,. \\  
\end{equation}
We find
\begin{align}\label{Hf}
H_{\rm f} &= \frac{\ii}{2} \left[(\vartheta_0^1)^\dagger \Gamma^7 \vartheta_0^2 - (\vartheta_0^2)^\dagger \Gamma^7 \vartheta_0^1\right] -\frac{3\ii}{4}\left[(\vartheta_0^1)^\dagger \Gamma^{78} \vartheta_0^1 - (\vartheta_0^2)^\dagger \Gamma^{78} \vartheta_0^2\right] \nn \\
&+ \sum_{n=1}^\infty \left[(\vartheta_{n}^1)^\dagger\left(\frac{\omega_n}{\alpha'p^+}-\frac{3\ii}{2}\Gamma^{78}\right)\vartheta_{n}^1 + (\vartheta_{n}^2)^\dagger\left(\frac{\omega_n}{\alpha'p^+}+\frac{3\ii}{2}\Gamma^{78}\right)\vartheta_{n}^2\right] + \varepsilon_{\rm f}\,,
\end{align}
where
\begin{equation}
\varepsilon_{\rm f} := -\frac{8}{\alpha' p^+} \sum_{n=1}^\infty \omega_n\,.
\end{equation}
In deriving the normal-ordering constant $\varepsilon_{\rm f}$ we have used ${\rm Tr}(P_\pm) = 8$ and ${\rm Tr}(P_\pm \Gamma^7) = {\rm Tr}(P_\pm \Gamma^{78}) = 0$.

\subsection{Light-cone Hamiltonian}
Combining the bosonic \eqref{Hb} and fermionic \eqref{Hf} parts we obtain the Hamiltonian
\begin{align}
H &= \frac{1}{\alpha' p^+} \sum_{I=1}^8 \left[ 
\omega^I_0 (\alpha_0^I)^\dagger \alpha^I_0 +
\sum_{n=1}^\infty \omega_n^I\left((\alpha_n^I)^\dagger \alpha^I_n + (\tilde{\alpha}_n^I)^\dagger \tilde{\alpha}^I_n \right)
\right] \nn \\
&+\frac{\ii}{2} \left[(\vartheta_0^1)^\dagger \Gamma^7 \vartheta_0^2 - (\vartheta_0^2)^\dagger \Gamma^7 \vartheta_0^1\right] -\frac{3\ii}{4}\left[(\vartheta_0^1)^\dagger \Gamma^{78} \vartheta_0^1 - (\vartheta_0^2)^\dagger \Gamma^{78} \vartheta_0^2\right] \nn \\
&+ \sum_{n=1}^\infty \left[(\vartheta_{n}^1)^\dagger\left(\frac{\omega_n}{\alpha'p^+}-\frac{3\ii}{2}\Gamma^{78}\right)\vartheta_{n}^1 + (\vartheta_{n}^2)^\dagger\left(\frac{\omega_n}{\alpha'p^+}+\frac{3\ii}{2}\Gamma^{78}\right)\vartheta_{n}^2\right] + \varepsilon\,,
\end{align}
where the normal-ordering constant $\varepsilon$ is the sum of the bosonic and fermionic ones, whose contributions cancel, up to a finite number of terms:
\begin{equation}
\varepsilon = \varepsilon_{\rm f} + \varepsilon_{\rm b} = 6 + \frac{2}{\alpha'p^+} \sum_{n=1}^{2\sqrt{2}\alpha'p^+}\left(3\alpha'p^+ - \omega_n\right)\,.
\end{equation}

\section*{Acknowledgments}
I would like to thank Niall Macpherson, Rodolfo Russo, Alessandro Tomasiello, Arkady Tseytlin and Kentaroh Yoshida for helpful communication. My work was funded, in whole, by ANR (Agence Nationale de la Recherche) of France under contract number ANR-22-ERCS-0009-01. 

\bibliography{draft}
\bibliographystyle{utphys}

\end{document}